\documentclass[aps,prd,showpacs,nofootinbib,preprint]{revtex4}

\usepackage{amsmath}
\usepackage{amssymb}
\usepackage{graphicx}
\usepackage{verbatim}

\begin{document}
\pagestyle{plain}

\title{Inflating wormholes in the braneworld models}
\author{K. C. Wong}
\email{fankywong@gmail.com}
\affiliation{Department of Physics
and Center for Theoretical and Computational Physics, University
of Hong Kong, Pok Fu Lam Road, Hong Kong, P. R. China}
\author{T. Harko}
\email{harko@hkucc.hku.hk}
\affiliation{Department of Physics and
Center for Theoretical and Computational Physics, University of
Hong Kong, Pok Fu Lam Road, Hong Kong, P. R. China}
\date{\today}
\author{K. S. Cheng}
\email{hrspksc@hkucc.hku.hk}
\affiliation{Department of Physics
and Center for Theoretical and Computational Physics, University
of Hong Kong, Pok Fu Lam Road, Hong Kong, P. R. China}

\begin{abstract}
The braneworld model, in which our Universe is a three-brane embedded
in a five-dimensional bulk, allows the existence of wormholes, without any violation of the energy conditions.
A fundamental ingredient of traversable wormholes is the violation of the null energy condition (NEC).
However, in the brane world models, the stress energy tensor confined on the brane, threading the wormhole, satisfies the NEC. In conventional general relativity, wormholes existing before inflation can be significantly enlarged by the expanding spacetime. We investigate the evolution of an inflating wormhole in the brane world scenario, in which the wormhole is supported by the nonlocal brane world effects. As a first step in our study we consider the possibility of embedding a four-dimensional brane world wormhole into a five dimensional bulk. The conditions for the embedding are obtained by studying the junction conditions for the wormhole geometry, as well as the full set of the five dimensional bulk field equations. For the description of the inflation we adopt the chaotic inflation model. We study the dynamics of the brane world wormholes during the exponential inflation stage, and in the stage of the oscillating scalar field. A particular exact solution corresponding to a zero redshift wormhole is also obtained. The resulting evolution shows that while the physical and geometrical parameters of a zero redshift wormhole decay naturally, a wormhole satisfying some very general initial conditions could turn into a black hole, and exist forever.
\end{abstract}


\maketitle
\section{Introduction}

Wormholes are hypothetical connected spacetime
\cite{Morris:1988cz,Visser}, and are primarily useful as
``gedanken-experiments'' and as a theoretician's probe of the
foundations of general relativity. The wormhole has no event horizon, which allows a two-way passage of energy or light signals. It is possible to transform wormholes into time machines for backward time travel \cite{Morris2,Frolov}. Wormhole solutions are a specific
example in solving the Einstein field equation in the reverse
direction, namely, one first considers an interesting space-time
metric, then one finds the matter source, responsible for the
respective geometry. In this manner, it was found that some of
these solutions possess a peculiar property, namely ``exotic
matter'', involving a stress-energy tensor that violates the null
energy condition.  A number of specific solutions with this property have been
found (\cite{solutions} and references therein). These
geometries also allow closed timelike curves, with the respective
causality violations \cite{Morris:1988tu}. However, the existence of wormholes without horizon would lead to the violation of the energy conditions at the throat of the wormholes \cite{Hawking,Tipler}. This implies that for some observers wormholes would require negative energy.

The conventional manner of finding wormhole solutions is
essentially to consider an interesting space-time metric, and then
to derive the stress-energy tensor components. A more systematic
approach in searching for exact solutions, namely, by assuming
spherical symmetry, and the existence of a non-static
conformal symmetry, was considered in \cite{Boehmer:2007rm}.

The requirement of negative energy makes the existence of a considerable number of wormholes nowadays very unlikely. However, the existence of wormholes in the early Universe seems to be more natural, as the quantum mechanical effects would be much stronger. Quantum mechanical effects may provide the negative energy, like, for example, via the Casimir effect. An inflating wormhole in conventional general relativity was already proposed and studied by  Roman \cite{Roman}. However, the systematic analysis of the cosmological implications of the existence of the  wormholes in the early Universe still remains an open problem.

While the negative energy required by a wormhole to exist is the main problem of the theory in standard general relativity, the braneworld models of the Universe provide a natural way for the existence of the wormholes.  In this work, we consider the Randall-Sandrum brane world model, which were originally
introduced to give an alternative to the compactification of the extra dimensions \cite{Randall,Randall2},
and to explain the hierarchy problem of the particle physics. Standard 4D gravity can be recovered in the low-energy limit of
the model, by assuming that a 3-brane of positive tension is embedded in a 5D
anti-de Sitter bulk. The covariant formulation of the brane world
models has been developed in \cite{Shiromizu}, leading to the
modification of the standard Friedmann equations on the brane. It
turns out that the dynamics of the early universe is altered by
the quadratic terms in the energy density and by the components of the bulk Weyl tensor, which both give a
contribution in the energy momentum tensor. This implies a
modification of the basic equations describing the cosmological
and astrophysical dynamics, which has been extensively considered
recently \cite{all2}. There is a number of works that use the brane world model to solve various cosmological problems. For instance, matter exchange between the brane and the bulk has been considered in \cite{Ap, Umezu}, and this gives a possible scenario of reheating \cite{Harko08}. Scalar field inflation \cite{Mizuno}, or non-scalar field inflation \cite{Wong}, have also been considered in the brane world models.

In a brane world Universe wormholes could
 exist without negative energy. This was pointed out by Lobo \cite{Lobo}, who showed that due to the braneworld corrections to the  energy-momentum tensor  a
 static wormhole in a RSII brane could obey the energy condition everywhere. The terms that allows the energy condition of the wormhole to be satisfied  in the brane world model come from the nonlocal projected Weyl tensor, and they appear as a correction terms in the $4D$ Einstein equation. The braneworld correction in energy is equivalent to  ordinary matter in conventional general relativity \cite{Leon}.

An interesting question, raised first in \cite{Ku08}, is the possibility of wormhole-black hole transition. It is quite likely, in view of some recent studies, that a time-dependent equation of state had caused the Universe to evolve from an earlier phantom-energy model. In that case traversable wormholes could have formed spontaneously. Such wormholes would eventually transform into black holes. This would provide a possible explanation for the existence of a huge number of black hole candidates, while any evidence for the existence of wormholes is entirely lacking.

It is the purpose of the present paper to consider the general properties and the dynamics of an inflating wormhole in the brane world scenario. As a first step in our study we consider the problem of the embedding of a braneworld wormhole into a five dimensional bulk. By considering a general five dimensional metric, with all metric tensor components function of time, radial coordinate and the extra-dimension, we formulate first the junctions conditions that must be satisfied by the wormhole geometry. Then, by considering the full system of five-dimensional Einstein field equations we obtain the five-dimensional equations satisfied by the wormhole metric tensor components. Particular cases of embedding (static geometry, homogeneous energy-momentum tensor and scalar field dominated braneworld models) are considered in detail, and it is shown that for each of these cases wormhole models can be obtained, at least in principle, in a full five-dimensional setup.  As a next step in our study we consider the evolution of an inflating (four-dimensional) braneworld wormhole in the early Universe. The inflation model we choose is the  chaotic inflation model in the RSII braneworld  model, considered in \cite{Liddle}. Current observations still permit the quadratic potential chaotic inflation model in the braneworld. On the contrary, the quadric and higher order exponents are excluded in the high energy regime. We study the evolution of a braneworld inflating wormhole in a chaotic inflation model, with quadratic potential, i.e. $V(\phi)=m^2\phi^2/2$, where $m$ is the mass of the inflaton field, and $\phi $ is the scalar field. We assume that during inflation a braneworld wormhole that does not violate the energy conditions is created by the scalar field, or by the local perturbations of the scalar field. We consider how such a wormhole would evolve as the Universe inflates, reheats, and becomes radiation dominant. Finally, we consider the conditions under which the transition of the wormhole to a black hole could have taken place in the early Universe.

The present paper is organized as follows. In Section~\ref{sect1} we consider the general problem of the embedding of a braneworld wormhole into the bulk geometry. Particular cases of embedding are considered in Section~\ref{sect2}. In Section~\ref{reviewbw}, we review, following \cite{Lobo},  the properties of the braneworld wormholes,  and show that they can exist without the violation of energy condition.  In Section~\ref{infbwh}, we introduce the inflating braneworld wormhole model. The basic equations describing the braneworld evolution of the inflating wormhole are presented in Section~\ref{4}. A simple solution of the field equations corresponding to a zero redshift wormhole, is obtained in Section~\ref{zeroredshift}. The chaotic inflation on the brane is reviewed in Section~\ref{6}. The evolution of the inflating wormhole during the exponential inflation and oscillating phases is analyzed in Section~\ref{soln}. We discuss and conclude our results in Section~\ref{astroimp}. In the present paper we use the natural system of units with $G=c=1$


\section{The bulk configuration of the braneworld wormholes}\label{sect1}

The four-dimensional wormhole models we are going to consider rely only on the non-local projected Weyl tensor of the bulk. Hence it would be important to  investigate what kind of bulk configuration can lead to a Weyl-tensor supported four-dimensional wormhole. The wormhole on the brane could possibly arise due to  a specific form of the Weyl tensor of the bulk, like, for example,  the one corresponding to a gravitational wave, or it may arise because our brane is located in such a way that the homogeneity is broken, so that the bulk metric is "abnormally" projected. Of course the specific form of the Weyl tensor supporting a four-dimensional wormhole can also be the combination of both previously mentioned effects. Therefore we can consider the full gauge freedom of the metric of the bulk, such that it projects as a wormhole on the brane, i.e., the metric coefficients  $g_{AB}$ on the bulk could all be non-zero. In this way we may think on the wormhole as an effect of a 5D gravitational wave. However, as motivated by the fact that an expanding Universe can be modeled as a motion of the brane in the bulk, we would like to model the brane wormhole by allowing the brane to offset its position.

In the following we denote by capital letters the coordinates in the bulk, with values running from $0$ to $4$. Greek letters mean coordinates on the brane, while lower case letters mean spatial coordinate. We denote by $n_A$ the normal vector to the brane, and by $y$ the fifth coordinate. The derivatives with respect to the coordinates are denoted by a comma.

\subsection{The induced metric and the junction conditions for braneworld wormholes}

In a suitable system of coordinates, the expansion of the Universe can be viewed as the motion of the brane through a static bulk \cite{Bowcock, Mukohyama}. Suppose a region of the brane slowly expands in a non-homogeneous manner. This creates a local inhomogeneity. However, we assume that asymptotically the expanding Universe is homogeneous. Due to its motion, the brane will now be located into a new position
\begin{equation}\label{coTrans}
y=Y(r),
\end{equation}
where we assume that $Y$ has no dependency on $t$. We assume that the 3-homogeneity of the bulk metric is broken locally. Hence the Gaussian coordinate system may not be the normal coordinate system locally.  Therefore for the bulk metric we assume the general form
\begin{equation}\label{bulk}
ds^2=-M(t,r,y)^2dt^2+N(t,r,y)^2dr^2+P(t,r,y)^2\left(d\theta^2+\sin^2\theta d\phi^2\right)+Q(t,r,y)^2dy^2,
\end{equation}
with all the metric coefficients functions of the time and radial distance coordinate, as well as of the five-dimensional coordinate $y$.
The metric tensor coefficients satisfy the bulk Einstein equation
\begin{equation}
{}^{(5)}G_{AB}=-^{(5)}\Lambda{}^{(5)}g_{AB}+k_5^2\left[{}^{(5)}T_{AB}+\delta(y-Y(r))T^{\rm brane}_{AB}\right],
\end{equation}
where $^{(5)}\Lambda$ is the five-dimensional cosmological constant, $^{(5)}T_{AB}$ is the bulk matter energy- momentum tensor, and $T^{\rm brane}_{AB}$ is the energy-momentum tensor of the matter on the brane, respectively. The five-dimensional gravitational coupling constant is denoted by $k_5^2$.

If we use $t,r,\theta,\phi$ as brane coordinates, the normal vector $n$ can no longer be a constant throughout the brane, since the brane is offset. The induced metric on the brane is obtained by substituting $y=Y(r)$ in the metric coefficients, and by taking into account the presence of $Qdy$ in the bulk, as \cite{Binetruy}
\begin{eqnarray}
{}^{(4)}ds^2_{induced}&=&-M(t,r,Y(r))^2dt^2+\left[N(t,r,Y(r))^2+Q(t,r,Y(r))^2Y_{,r}(r)^2\right]dr^2+\nonumber\\
&&P(t,r,Y(r))^2\left(d\theta^2+\sin^2\theta d\phi^2\right).
\end{eqnarray}
In particular, this metric will induce an inflating wormhole if the metric coefficients satisfy the following conditions
\begin{eqnarray}
M(t,r,Y(r))^2&=&e^{2\Phi(r,t)},\label{bulkcon0}\\
N(t,r,Y(r))^2+Q(t,r,Y(r))^2Y_{,r}(r)^2&=&\frac{a(t)^2}{1-b(r)/r},\label{bulkcon1}\\
P(t,r,Y(r))^2&=&a(t)^2r^2.\label{bulkcon2}
\end{eqnarray}

$e^{2\Phi(r,t)}$ is called the redshift function, and $b(r)$ is called the throat function of the wormhole. The throat of a wormhole is defined by $b(r_0)=r_0$. A time dependent throat function would lead to divergence of energy density at the throat and therefore we do not consider it. The function $a(t)$ is the scale factor describing the expansion of the wormhole.

By integrating the bulk Einstein equations along the extra dimension it follows that the matter content in the brane affects the bulk metric through the jump of the brane extrinsic curvature $K_{\mu\nu}$ \cite{Israel}. This is the junction condition of the brane. The extrinsic curvature can be calculated according to
\begin{equation}
K_{AB}=h_A^C\nabla_Cn_B,
\end{equation}
where
\begin{equation}
h_{AB}=g_{AB}-n_{A}n_{B},
\end{equation}
is the projection tensor which is a tensor in the bulk that acts as metric on the brane. The brane is assumed to have the same properties on both sides. Mathematically this represents the $\mathcal{Z}_2$ symmetry. By imposing the $\mathcal{Z}_2$ symmetry we can relate the extrinsic curvature to the energy-momentum tensor of the matter on the brane,
\begin{equation}\label{junction}
K_{\mu\nu}(r,t,R(y^+))-K_{\mu\nu}(r,t,R(y^-))=-2K_{\mu\nu}(r,t,R(y))=-k_5^2\left(T_{\mu\nu}-\frac{1}{3}Th_{\mu\nu}\right),
\end{equation}
where the second equality follow from the $\mathcal{Z}_2$ symmetry of the brane, and the metric and the energy-momentum tensors of the brane can be obtained from the corresponding tensors of the bulk by using the coordinate transformation defined by Eq.~(\ref{coTrans}). The extrinsic curvature specifies the derivative of the metric along the flow defined by the normal vector. It tells us how the metric evolves off the brane in order to match the external space. It also sets the boundary conditions for off brane evolution. The components of the normal vector can be calculated according to
\begin{equation}
n_A=\frac{\xi_A}{\sqrt{\xi^C\xi_C}},
\end{equation}
where
\begin{equation}
\xi_A=\nabla_A\left(y-Y(r)\right),
\end{equation}
which give
\begin{equation}
n_A=\left[0,-\frac{Y_{,r}}{\sqrt{1/Q^2+Y_{,r}^2/N^2}},0,0,\frac{1}{\sqrt{1/Q^2+Y_{,r}^2/N^2}}\right].
\end{equation}
The energy - momentum tensor of the matter on the brane is given by a delta function $\delta(y-Y(r))T^{\rm brane}_{AB}$. For the brane energy - momentum tensor $T^{\rm brane}_{AB}$ we assume the perfect fluid form,
\begin{equation}\label{TbraneInBulk}
T^{\rm brane}_{AB}=\left[\rho(t,r)+p(t,r)\right]U_{A}U_{B}+p(t,r)h_{AB},
\end{equation}
with $\rho(t,r)$ and $p(t,r)$ are the energy density and the pressure of the matter on the brane, respectively. $U_{A}$ is the five - dimensional velocity field of the matter fluid on the brane. $T^{\rm brane}_{AB}$ satisfies the condition $T^{\rm brane}_{AB}n^A=0$.  The four-dimensional energy-momentum tensor $T_{\mu\nu}$ can be obtained from $T^{\rm brane}_{AB}$   by the relation $T_{\mu\nu}=e^{A}_{\mu}e^{B}_{\nu}T_{AB}$, where $e^{\mu}$ are the tetrad vectors that form an orthogonal basis on the brane \cite{Binetruy}. In the following we assume that the fluid is at rest with respect to the comoving observers, so that $U_{A}=[e^{\Phi},0,0,0,0]$. Therefore the junction conditions can be written as
\begin{equation}\label{k00}
K_{tt}=-\frac{M\left(Y_{,r}M_{,r}Q^2-M_{,y}N^2\right)}{NQ\sqrt{N^2+Y_{,r}^2Q^2}}=-\frac{k_5^2(2\rho+3p)M^2}{6},
\end{equation}
\begin{equation}\label{k01}
K_{tr}=\frac{-Y_{,r}\left(Q_{,t}N-N_{,t}Q\right)}{N^2+Y_{,r}^2Q^2}=0,
\end{equation}
\begin{eqnarray}\label{k11}
K_{rr}&=&\frac{N(N^2+2Y_{,r}^2Q^2)}{Q(N^2+Y_{,r}^2Q^2)^{5/2}}\times \nonumber\\
&&\left[NY_{,r}N_{,r}Q^2+\left(N^3+2NY^2Y_{,r}^2Q^2\right)N_{,y}-\left(2Y_{,r}N^2Q+Y_{,r}^3Q^3\right)Q_{,r}\right.-\nonumber\\
&&\left.Y_{,r,r}Q^2N^2-Y_{,r}^2QQ_{,y}N^2\right]=
\frac{k_5^2\rho}{6}\frac{N^2(N^2+2Y_{,r}^2Q^2)}{N^2+Y_{,r}^2Q^2},
\end{eqnarray}
\begin{equation}\label{k22}
K_{\theta\theta}=\frac{P\left(-Y_{,r}P_{,r}Q^2+P_{,y}N^2\right)}{NQ(N^2+Y_{,r}^2Q^2)^{1/2}}=\frac{k_5^2}{6}P^2\rho ,
\end{equation}
\begin{equation}
K_{\phi\phi}=\sin^2\theta K_{\theta\theta}.
\end{equation}
In order for a brane-world wormhole be embedded into the bulk these junction conditions must be satisfied. The first condition for the embedding is obtained from Eq.~(\ref{k01}) as
\begin{equation}\label{bulkcon4}
\frac{N_{,t}(t,r,Y(r))}{N(r,t,Y(r)}=\frac{Q_{,t}(t,r,Y(r))}{Q(r,t,Y(r))},
\end{equation}
since we have assumed that $Y_{,r}\neq 0$. In a brane without radial energy flow, a co-expanding wormhole that travels in the $y$ direction imposes the condition that the metric coefficients $N$ and $Q$ have a similar time behavior on the brane. By substituting Eq.~(\ref{bulkcon4}) into the time derivative of Eq.~(\ref{bulkcon1}) we obtain a separable equation with solutions of the form
\begin{equation}\begin{array}{ccc}\label{NQseparate}
N(t,r,Y(r))=a(t)\mathcal{N}(r), Q(t,r,Y(r))=a(t)\mathcal{Q}(r),
\end{array}\end{equation}
where $\mathcal{N}(r)$ and $\mathcal{Q}(r)$ are some arbitrary functions of the radial coordinate only.
The position of the brane $Y(r)$ on which an inflating wormhole can exist is given by the equation
\begin{equation}\label{Yr}
Y_{,r}^2(r)=\frac{1}{\mathcal{Q}^2}\left[\frac{1}{1-b(r)/r}-\mathcal{N}^2\right].
\end{equation}
The throat of the wormhole can be pictured as a travel upward relative to the brane, i.e., $Y'(r_0)\rightarrow\infty$.
Due to the Codazzi equation for the extrinsic curvature $\nabla_BK^B_A-\nabla_AK={}^{(5)}R_{BC}g_A^Bn^c$, the junction condition also implies the local conservation of the energy-momentum tensor. On a source free bulk we obtain the standard conservation equation of the energy-momentum tensor on the brane,
\begin{equation}\label{localconservation}
{}^{(4)}\nabla^{\mu}{}^{(4)}T_{\mu\nu}=0,
\end{equation}
which gives the following constraints on $a$ and $\Phi$ with respect to the evolution of the matter on the brane,
\begin{equation}
\rho_{,t}(t,r)+3\frac{\dot{a}}{a}\left[\rho(t,r)+p(t,r)\right]=0,\label{branecon1}
\end{equation}
\begin{equation}
\partial_{,r}\Phi \left[\rho(t,r)+p(t,r)\right]+\partial_r p(t,r)=0.\label{branecon2}
\end{equation}
In the above equations we have used the brane metric only.

\subsection{The gravitational field equations}

The final, but the most important constraints on the metric follow from the  $5D$ Einstein equations. The independent components of the Einstein tensor of the metric given by Eq.~(\ref{bulk}) are given by
\begin{eqnarray}\label{G00}
G_{00}&=&\left(\frac{2M^2P_{,r}}{PN^3}+\frac{Q_{,r}M^2}{QN^3}\right)N_{,r}+%
\left(\frac{2P_{,t}}{PN}+\frac{Q_{,t}}{NQ}\right)N_{,t}%
+\left(-\frac{2M^2P_{,y}}{PNQ^2}+\frac{Q_{,y}M^2}{NQ^3}\right)N_{,y}-\nonumber\\
&&\left(\frac{P_{,r}M^2}{P^2N^2}+\frac{2Q_{,r}M^2}{PN^2Q}\right)P_{,r}+\left(\frac{P_{,t}}{P}+\frac{2Q_{,t}}{PQ}\right)P_{,t}%
+\left(\frac{2Q_{,y}M^2}{PQ^3}-\frac{P_{,y}M^2}{P^2Q^2}\right)P_{,y}-\nonumber\\
&&\frac{N_{,y,y}M^2}{NQ^2}-\frac{2P_{,r,r}M^2}{PN^2}
-\frac{2P_{,y,y}M^2}{PQ^2}-\frac{Q_{,r,r}M^2}{N^2Q}+\frac{M^2}{P^2},
\end{eqnarray}
\begin{equation}\label{G01}
G_{01}=\left(\frac{2P_{,t}}{PM}+\frac{Q_{,t}}{MQ}\right)M_{,r}+\left(\frac{2P_{,r}}{PN}+\frac{Q_{,r}}{NQ}\right)N_{,t}%
-\frac{2P_{,r,t}}{P}-\frac{Q_{,r,t}}{Q},
\end{equation}
\begin{equation}\label{G05}
G_{04}=\left(\frac{N_t}{MN}+\frac{2P_{,t}}{PM}\right)M_{,y}+\left(\frac{N_{,y}}{NQ}+\frac{2P_{,y}}{PQ}\right)Q_{,t}%
-\frac{N_{,t,y}}{N}-\frac{2P_{,t,y}}{P},
\end{equation}
\begin{eqnarray}\label{G11}
G_{11}&=&\left(\frac{2P_{,r}}{PM}+\frac{Q_{,r}}{MQ}\right)M_{,r}+\left(\frac{2N^2P_{,t}}{PM^3}+\frac{Q_{,t}N^2}{QM^3}\right)M_{,t}%
+\left(\frac{2N^2P_{,y}}{MPQ^2}-\frac{Q_{,y}N^2}{MQ^3}\right)M_{,y}+\nonumber\\
&&\left(\frac{P_{,r}}{P}+\frac{2Q_{,r}}{PQ}\right)P_{,r}
-\left(\frac{P_{,t}N^2}{P^2M^2}+\frac{2Q_{,t}N^2}{PM^2Q}\right)P_{,t}+\left(\frac{P_{,y}N^2}{P^2Q^2}-\frac{2Q_{,y}N^2}{PQ^3}\right)P_{,y}+\nonumber\\
&&\frac{M_{,y,y}N^2}{MQ^2}-\frac{2P_{,t,t}N^2}{PM^2}+\frac{2P_{,y,y}N^2}{PQ^2}-\frac{Q_{,t,t}N^2}{M^2Q}-\frac{N^2}{P^2},
\end{eqnarray}
\begin{equation}\label{G15}
G_{14}=-\frac{M_{,r,y}}{M}+\frac{N_{,y}M_{,r}}{MN}+\frac{Q_{,r}M_{,y}}{MQ}-\frac{2P_{,r,y}}{P}+\frac{2N_{,y}P_{,r}}{PN}+\frac{2Q_{,r}P_{,y}}{PQ},
\end{equation}
\begin{eqnarray}\label{G22}
G_{22}&=&\left(-\frac{P^2N_{,r}}{MN^3}+\frac{PP_{,r}}{MN^2}+\frac{P^2Q_{,r}}{MN^2Q}\right)M_{,r}+\left(\frac{P^2N_{,t}}{NM^3}%
+\frac{PP_{,t}}{M^3}+\frac{P^2Q_{,t}}{QM^3}\right)M_{,t}+\nonumber\\
&&+\left(\frac{P^2N_{,y}}{MNQ^2}+\frac{PP_{,y}}{MQ^2}-\frac{Q_{,y}P^2}{MQ^3}\right)M_{,y}%
-\left(\frac{PP_{,r}}{N^3}+\frac{Q_{,r}P^2}{QN^3}\right)N_{,r}-\left(\frac{PP_{,t}}{NM^2}+\frac{Q_{,t}P^2}{NM^2Q}\right)N_{,t}+\nonumber\\
&&\left(\frac{PP_{,y}}{NQ^2}-\frac{Q_{,y}P^2}{NQ^3}\right)N_{,y}+\frac{PQ_{,r}P_{,r}}{N^2Q}-\frac{PQ_{,t}P_{,t}}{M^2Q}%
-\frac{PQ_{,y}P_{,y}}{Q^3}+\frac{P^2M_{,r,r}}{MN^2}+\frac{P^2M_{,y,y}}{MQ^2}-\nonumber\\
&&\frac{P^2N_{,t,t}}{M^2N}+\frac{P^2N_{,y,y}}{NQ^2}+\frac{PP_{,r,r}}{N^2}-\frac{PP_{,t,t}}{M^2}+\frac{PP_{,y,y}}{Q^2}%
+\frac{P^2Q_{,r,r}}{N^2Q}-\frac{P^2Q_{,t,t}}{M^2Q},
\end{eqnarray}
\begin{eqnarray}\label{G44}
G_{44}&=&\left(-\frac{Q^2N_{,r}}{MN^3}+\frac{2P_{,r}Q^2}{PMN^2}\right)M_{,r}+\left(\frac{Q^2N_{,t}}{NM^3}+\frac{2P_{,t}Q^2}{PM^3}\right)M_{,t}%
+\left(\frac{N_{,y}}{MN}+\frac{2P_{,y}}{PM}\right)M_{,y}+\nonumber\\
&&\left(\frac{P_{,r}Q^2}{P^2N^2}-\frac{2N_{,r}Q^2}{PN^3}\right)P_{,r}-\left(\frac{P_{,t}Q^2}{P^2M^2}+\frac{2N_{,t}Q^2}{PM^2N}\right)P_{,t}%
+\left(\frac{2N_{,y}}{PN}+\frac{P_{,y}}{P}\right)P_{,y}+\nonumber\\
&&\frac{M_{,r,r}Q^2}{MN^2}-\frac{N_{,t,t}Q^2}{M^2N}+
\frac{2P_{,r,r}Q^2}{PN^2}-\frac{2P_{,t,t}Q^2}{PM^2}-\frac{Q^2}{P^2}.
\end{eqnarray}
The energy momentum tensor of the bulk is specified by Eq.~(\ref{TbraneInBulk}),
\begin{equation}
T_{AB}=\left(\begin{matrix}
M^2\rho & 0 & 0 & 0 & 0 \\
0 & \frac{N^4p}{N^2+Y_{,r}^2Q^2} & 0 & 0 & \frac{Q^2NY_{,r}p}{N^2+Y_{,r}^2Q^2}\\
0 & 0 & P^2p & 0 & 0\\
0 & 0 & 0 & P^2\sin^2\theta p & 0\\
0 & \frac{Q^2NY_{,r}p}{N^2+Y_{,r}^2Q^2} & 0 & 0 & \frac{Q^4Y_{,r}^2p}{N^2+Y_{,r}^2Q^2}
\end{matrix}\right)_.
\end{equation}
In a source free bulk, two intermediate constraints of the metric can be obtained by setting $G_{01}$ and $G_{04}$ to zero,
\begin{equation}\label{bulkextra1}
\left(\frac{2P_{,t}}{PM}+\frac{Q_{,t}}{MQ}\right)M_{,r}+\left(\frac{2P_{,r}}{PN}+\frac{Q_{,r}}{NQ}\right)N_{,t}%
-\frac{2P_{,r,t}}{P}-\frac{Q_{,r,t}}{Q}=0,
\end{equation}
\begin{equation}\label{bulkextra2}
\left(\frac{N_{,t}}{MN}+\frac{2P_{,t}}{PM}\right)M_{,y}+\left(\frac{N_{,y}}{NQ}+\frac{2P_{,y}}{PQ}\right)Q_{,t}%
-\frac{N_{,t,y}}{N}-\frac{2P_{,t,y}}{P}=0.
\end{equation}
By differentiating with respect to $r$ both sides of Eq.~(\ref{bulkcon0}) gives
\begin{equation}\label{Mr}
M_{,r}+Y_{,r}M_{,y}=e^{\Phi}\partial_r\Phi.
\end{equation}
By eliminating $M_{,r}$ with the use of Eq.~(\ref{Mr}) from the junction condition Eq.~(\ref{k00}) we obtain
\begin{equation}\label{eqtMy}
M_{,y}=-\frac{k_5^2(2\rho+3p)MNQ}{6\sqrt{N^2+Y_{,r}^2Q^2}}+\frac{Y_{,r}e^{\Phi}\partial_r\Phi}{N^2+Y_{,r}^2Q^2}.
\end{equation}
>From Eq.~(\ref{bulkcon2}) we obtain
\begin{align}\label{decomP}
P_{,t}=\dot{a}r\mbox{ and }P_{,r}=a-Y_{,r}P_{,y},
\end{align}
while from Eq.~(\ref{NQseparate}) it follows that
\begin{equation}\label{decomN}
N_{,t}=\dot{a}\mathcal{N},\;N_{,r}=a\mathcal{N}_{,r}-Y_{,r}N_{,y},
\end{equation}
and
\begin{equation}\label{decomQ}
Q_{,t}=\dot{a}\mathcal{Q},\;Q_{,r}=a\mathcal{Q}_{,r}-Y_{,r}Q_{,y},
\end{equation}
respectively. By substituting the above relations and Eqs.~(\ref{bulkcon0}), (\ref{bulkcon1}), and (\ref{bulkcon2}) into Eq.~(\ref{bulkextra1}), we obtain the following  equation for the metric tensor coefficients in the bulk evaluated on the brane,
\begin{equation}\label{bulkextra1b}
\frac{\dot{a}}{a}\left[3\left(\partial_r\Phi-Y_{,r}\frac{M_{,y}}{M}\right)-\frac{Y_{,r}P_{,y}}{ar}-\frac{Y_{,r}Q_{,y}}{Q}\right]+\frac{Y_{,r}Q_{,y,t}}{Q}=0.
\end{equation}
By substituting Eq.~(\ref{decomP}) into Eq.~(\ref{k22}) gives the following partial differential equation for the metric coefficient $P$,
\begin{equation}\label{eqtPy}
P_{,y}=\frac{k_5^2\rho}{6}\frac{PNQ}{\sqrt{N^2+Y_{,r}^2Q^2}}+\frac{aY_{,r}Q^2}{N^2+Y_{,r}^2Q^2}.
\end{equation}

\section{Particular cases of embedding of brane world wormholes}\label{sect2}

In the previous Section we have formulated the basic equations describing the embedding of a four-dimensional braneworld wormhole into a five dimensional bulk. In the present Section we consider some particular cases of embedding.

\subsection{Braneworlds  with homogeneous energy - momentum tensor}

 As a first case in our study we consider the special case in which the energy and the pressure of the matter on the brane  does not depend on the radial coordinate $r$, and it is a function of time only. Then the local conservation Eq.~(\ref{branecon2}) requires $p=-\rho$ or $\partial_r\Phi=0$ on the brane.

\subsubsection{$p\neq-\rho$}

In this case, since $\partial_r\Phi=0$, Eq.~(\ref{Mr}) becomes a first order linear partial differential equation,
\begin{equation}
M_{,r}+Y_{,r}M_{,y}=0,
\end{equation}
with the general solution given by
\begin{equation}
M(t,r,y)=\mathcal{M}(t,y-Y(r)),
\end{equation}
where $\mathcal{M}(t,z)=\mathcal{M}(t,y-Y(r))$ is an arbitrary function of the arguments.
Eq.~(\ref{eqtMy}) becomes
\begin{equation}\label{homo1M}
\mathcal{M}_{,z}(t,y-Y(r))=-\frac{k_5^2aM}{6}\left(2\rho+3p\right)\frac{\mathcal{N}\mathcal{Q}}{\sqrt{\mathcal{N}^2+\mathcal{Q}^2Y_{,r}^2}}.
\end{equation}
Unless $2\rho+3p=0$, this equation implies that the following expression is a constant,  which for simplicity is taken to be $1$,
\begin{equation}
\frac{\mathcal{N}\mathcal{Q}}{\sqrt{\mathcal{N}^2+\mathcal{Q}^2Y_{,r}^2}}=1.
\end{equation}
In terms of the throat function $b(r)$ the above equation can be written as
\begin{equation}
\mathcal{N}\mathcal{Q}=\frac{1}{\sqrt{1-b(r)/r}}.
\end{equation}
Independently of the choice of $Y(r)$, the bulk metric diverges at the throat. Since $\mathcal{N}$ and $\mathcal{Q}$ are arbitrary integration functions, we can choose $\mathcal{Q}=1$, thus obtaining
$$\begin{array}{ccc}
\mathcal{N}^2=\frac{1}{1-b/r}, &\mbox{and}& Y_{,r}=0,
\end{array}$$
Using Eq.~(\ref{eqtPy}), we find
\begin{equation}
P_{,y}=\frac{k_5^2a^2r\rho}{6}\sqrt{1-b(r)/r}.
\end{equation}
>From Eq.~(\ref{bulkextra1b}) we can obtain $Q_{,y}$, while Eq.~(\ref{k11}) determines $N_{,y}$. The time and extra-dimensional evolution of the bulk geometry is described by Eq.~(\ref{bulkextra2}). Therefore the boundary conditions for the off brane evolution of the wormhole are completely determined.

\subsubsection{$p=-\rho$}

The local matter conservation given by Eq.~(\ref{branecon1}) tells us that if $p=-\rho$ then $\rho_{,t}=0$. Hence Eq.~ (\ref{branecon2}) is automatically satisfied. The freedom for the choice of the bulk metric for this type of matter is larger. For example, a realizable setup is given by the following brane geometry,
\begin{equation}
Y_{,r}^2=\frac{1}{1-b(r)/r},
\end{equation}
which means that the wormhole was created solely by the embedding, and
\begin{equation}
\mathcal{N}=\mathcal{Q}=1,
\end{equation}
respectively. Eqs.~(\ref{decomN}) and (\ref{decomQ}) simplify the junction condition given by Eq.~(\ref{k11}) into
\begin{equation}
Q_{,y}+N_{,y}=\frac{k_5^2\rho a^2}{6}\sqrt{1-\frac{b}{r}}+\frac{a(b'r-b)}{2r^2(1-b/r)^{1/2}},
\end{equation}
$M_{,y}$ and $P_{,y}$ can be written in a simple form as
\begin{equation}
M_{,y}=\left(\frac{k_5^2\rho a}{6}+\partial_r\Phi\right)e^{\Phi}\sqrt{1-\frac{b}{r}},
\end{equation}
\begin{equation}
P_{,y}=\left(\frac{k_5^2\rho a^2r}{6}\right)\sqrt{1-\frac{b}{r}}.
\end{equation}
 Once $M$ and $P_{,y}$ are given, the bulk Einstein equation Eq.~(\ref{bulkextra1b}) determines $Q_{,y}$. From Eq.~(\ref{bulkextra2}) the form of $M$ can be determined. This completely solves the problem of the embedding of the brane world wormhole into the bulk.

\subsection{Wormholes on a static brane}

Static brane world wormholes have already been studied in \cite{Lobo}. We would like now to investigate what would be the embedding for this static wormhole supported by the off-set of a brane. The induced metric is static,  corresponding to $a=1$. Since $\rho$ and $p$ are independent of time, the junction conditions given by Eqs.~(\ref{k00}) - (\ref{k22}) tell us that the bulk metric is also static. The local matter conservation given by Eq. (\ref{branecon1}) and Eqs.~(\ref{bulkextra1}) - (\ref{bulkextra2}) is also automatically satisfied. Eq.~(\ref{branecon2}) becomes
\begin{equation}
p'=-(\rho+p)\Phi'.
\end{equation}
A static wormhole is compatible with $Q=1$. Using the junction conditions we can work out the $y$ derivative of the metric coefficient as follows
\begin{eqnarray}
M_{,y}&=&\frac{Y_{,r}M_{,r}}{N^2}-\frac{k_5^2}{6}\left(2\rho+3p\right)\sqrt{1+Y_{,r}^2/N^2}M,\\
N_{,y}&=&\frac{Y_{,r}N_{,r}}{N^2+2Y_{,r}^2}+\frac{k_5^2N\rho}{6}\frac{(1+Y_{,r}^2/N^2)^{3/2}}{N^2+2Y_{,r}^2}+\frac{Y_{,r,r}N}{N^2+2Y_{,r}^2},\\
P_{,y}&=&\frac{Y_{,r}P_{,r}}{N^2}+\frac{k_5^2}{6}\rho\sqrt{1+Y_{,r}^2/N^2}.
\end{eqnarray}
In particular a vacuum brane solution with $\rho =p =0$ is given by
\begin{eqnarray}
M_{,y}&=&\frac{Y_{,r}M_{,r}}{N^2},\\
N_{,y}&=&\frac{Y_{,r}N_{,r}+Y_{,r,r}N}{N^2+2Y_{,r}^2},\\
P_{,y}&=&\frac{Y_{,r}P_{,r}}{N^2}.
\end{eqnarray}
Using Eq.~(\ref{decomP}) we obtain
\begin{equation}
P_{,y}=Y_{,r}\left[1-\frac{b(r)}{r}\right].
\end{equation}
 With the use of these equations in the $G_{44}$ and $G_{14}$ components of the bulk Einstein field equations, and limiting ourselves to the brane, one could obtain two second order non-linear ordinary differential equations for $M$ and $N$, which involve only the second derivatives with respect to $r$ of the metric functions. Therefore a consistent solution for a wormhole for a static brane always exists.

\subsection{Wormholes on a scalar field filled brane}

The energy momentum tensor of a scalar field is given by
\begin{equation}\label{scalarfieldT}
T_{\mu\nu}=\partial_{\mu}\phi\partial_{\nu}\phi- g_{\mu\nu}\left[\frac{1}{2}g^{\alpha\beta}\partial_{\alpha}\phi\partial_{\beta}\phi+V(\phi)\right],
\end{equation}
where $V\left(\phi \right)$ is the self-interaction potential of the field.
In the space-time described by the metric given by Eqs.~(\ref{bulkcon0})-(\ref{bulkcon2}), the brane energy momentum tensor of the scalar field is given by
\begin{eqnarray}
T_{\mu\nu}&=&{\rm diag}\left\{e^{2\Phi}\left[e^{-2\Phi}\frac{1}{2}\dot{\phi}^2+V(\phi)\right], \frac{a^2r}{r-b}\left[e^{-2\Phi}\frac{1}{2}\dot{\phi}^2-V(\phi)\right]\right.,\nonumber\\ &&\left.a^2r^2\left[e^{-2\Phi}\frac{1}{2}\dot{\phi}^2-V(\phi)\right], a^2r^2\sin^2\theta\left[e^{-2\Phi}\frac{1}{2}\dot{\phi}^2-V(\phi)\right]\right\}.
\end{eqnarray}
This form of the energy-momentum tensor allows us to introduce an effective energy density and pressure according to the definitions
\begin{equation}\label{energyasso}
\rho(t,r)=\frac{1}{2}\dot{\phi(t)}^2e^{-2\Phi(r,t)}+V(\phi(t)), \; p(t,r)=\frac{1}{2}\dot{\phi(t)}^2e^{-2\Phi(r,t)}-V(\phi(t)).
\end{equation}
 With this forms of the effective energy density and pressure for the scalar field the conservation equation Eq.~(\ref{branecon2}) is automatically satisfied. If the scalar field is potential dominated, it satisfies an  effective equation of state  $p=-\rho$.  Moreover, the evolution of the scalar field is governed by the Klein - Gordon equation
\begin{equation}
-\frac{1}{\sqrt{-g}}\partial_{\mu}\left(\sqrt{-g}g^{\mu\nu}\partial_{\nu}\phi\right)+V'\left(\phi \right)=0,
\end{equation}
where $\sqrt{-g}$ is the square root of the determinant of the metric tensor. In the geometry of Eqs.~(\ref{bulkcon0})-(\ref{bulkcon2}), the evolution equation of the scalar field becomes
\begin{equation}\label{con1}
\ddot{\phi}+\left(3\frac{\dot{a}}{a}-\frac{\dot{M}}{M}\right)\dot{\phi}+V'(\phi)M^2=0,
\end{equation}
which is consistent with the substitution of  Eqs.~(\ref{energyasso}) into the conservation equation Eq.~(\ref{branecon1}). Since asymptotically the wormhole geometry tends to the global (cosmological) geometry of the spacetime, by taking the limit $r\rightarrow\infty$ in Eq.~(\ref{con1}), we found that the evolution of the scalar field $\phi $ results from the global cosmological evolution,
\begin{equation}
\ddot{\phi}+3\frac{\dot{a}}{a}\dot{\phi}+V'(\phi)=0.
\end{equation}
Therefore we obtain
\begin{equation}\label{bulkcon3}
\frac{M_{,t}(t,r,Y(r))}{M(t,r,Y(r))}=\frac{V'(\phi)}{\dot{\phi}}\left[M^2(t,r,Y(r))-1\right].
\end{equation}
This equation can be integrated by using the substitution $x=M^2$, and its general solution can be written as
\begin{equation}
M(t,r,Y(r))=\sqrt{\frac{1}{1-\mathcal{M}(r)\mathcal{A}(t)}},
\end{equation}
where
\begin{equation}
\mathcal{A}(t)=e^{\int{\frac{2V'(\phi)}{\dot{\phi}}}dt},
\end{equation}
and $\mathcal{M}(r)$ is an arbitrary integration function. Using Eqs.~(\ref{bulkextra2}) and (\ref{bulkextra1b})
a consistent set of $M$, $N$, $P$ and $Q$ can be obtained. Therefore,  an inflating brane world wormhole that does not perturb the local evolution of the cosmological scalar field does exist.
However,  instead of studying its details by solving the close system of bulk Einstein equation,
in the following Sections we derive some important results on brane world wormholes by using the effective brane description of the system.

\section{The brane geometry of static wormholes}\label{reviewbw}

In the covariant formulation of the RSII brane world model, the effective Einstein equation on the 4D brane is given by \cite{Shiromizu}
\begin{equation}\label{einstein}
G_{\mu\nu}=8\pi\left(T_{\mu\nu}+\frac{6}{\lambda}S_{\mu\nu}\right)-\varepsilon_{\mu\nu},
\end{equation}
where $\lambda $ is the brane tension, $T_{\mu\nu}$ is the energy momentum tensor of the matter on the brane, and
\begin{equation}
S_{\mu \nu }=\frac{1}{2}TT_{\mu \nu }-\frac{1}{4}T_{\mu \alpha
}T_{\nu }^{\alpha }+\frac{3T_{\alpha \beta }T^{\alpha \beta
}-T^{2}}{24}g_{\mu \nu },
\end{equation}
is a quadratic term in the energy momentum tensor,  which follows from the junction conditions of the embedding of the brane to the bulk. $T=T_{\mu}^{\mu }$ is the trace of the energy-momentum tensor. On the other hand, $\varepsilon_{\mu\nu}=C_{ABCD}n^{C}n^{D}g_{\mu }^{A}g_{\nu
}^{B}$ is  the projection of the bulk Weyl tensor $C_{ABCD}$ to the brane.
The metric $g^{(5)}_{AB}$ on the bulk induce a metric $g^{(4)}_{\mu\nu}$ on the brane by the map that embed the brane. The metric of a static wormhole on the brane is given by \cite{Lobo}
\begin{equation}
ds^2=-e^{2\Phi(r)}dt^2+\frac{dr^2}{1-b(r)/r}+r^2(d\theta^2+\sin^2\theta d^2\phi),\label{lobometric}
\end{equation}
where $\Phi(r)$ and $b(r)$ are called the redshift function and the form function, respectively.
To obtain a wormhole solution, several properties need to
be imposed, namely \cite{Morris:1988cz}: The throat is located at $r = r_0$
and $b\left(r_0\right) = r_0$. A flaring out condition of the throat is
required, i. e., $\left(b-b'r\right)/b^2 > 0$, where the prime denotes the derivative with respect to $r$. At the throat this inequality reduces to $b'\left(r_0\right) < 1$. The condition $1-b/r \geq 0$ is also required.
To be traversable,  there must be no
horizons present, which are identified as the surfaces with
$e^{2\Phi}\rightarrow 0$. Therefore $\Phi (r)$ must be finite everywhere.
For this metric the Einstein tensor $G_{\mu\nu}=R_{\mu\nu}-\frac{1}{2}g_{\mu\nu}R$  is
\begin{eqnarray}
G_{\mu\nu}&=&{\rm diag}\left\{\frac{e^{2\Phi}b'}{r^2}, \frac{r}{r-b}\left[\frac{2\Phi'}{r}\bigl(1-\frac{b}{r}\bigr)-\frac{b}{r^3}\right], r^2\Bigl[(\Phi''+\Phi'^2)\left(1-\frac{b}{r}\right)+\frac{\Phi'}{r}\left(1-\frac{b}{2r}-\frac{b'}{2}\right)\right.\nonumber\\
&&\left.+\frac{1}{2r^2}\left(\frac{b}{r}-b'\right)\Bigr], r^2\sin^2\theta\Bigl[(\Phi''+\Phi'^2)\left(1-\frac{b}{r}\right)\right.+\nonumber\\
&&\left.\frac{\Phi'}{r}\left(1-\frac{b}{2r} -\frac{b'}{2}\right)+\frac{1}{2r^2}\left(\frac{b}{r}-b'\right)\Bigr]\right\}.
\end{eqnarray}
The field equations on the brane can be written as $G_{\mu \nu}=8\pi T_{\mu \nu }^{{\rm eff}}$, where $T_{\mu \nu }^{{\rm eff}}=T_{\mu \nu}-\left(1/8\pi \right)\varepsilon _{\mu \nu}+\left(6/\lambda \right)S_{\mu \nu}$. Once  the energy momentum tensor on the brane $T_{\mu\nu}$ is known, we can obtain the projected Weyl tensor by $\varepsilon_{\mu\nu}=8\pi T^{\rm{mat}}_{\mu\nu}-G_{\mu\nu}$, where $T^{\rm{mat}}_{\mu\nu}=T_{\mu\nu}+(6/\lambda)S_{\mu\nu}$. Since $\varepsilon_{\mu\nu}$ is traceless, taking the trace of the brane Einstein equation Eq.~(\ref{einstein}) would give the constraint equation $8\pi {\rm Tr}T^{\rm{mat}}_{\mu\nu}={\rm Tr}G_{\mu\nu}=-R$, which relates the wormhole redshift and form functions with the matter component. From the trace of the Einstein tensor we obtain
\begin{equation}
R=-\left(1-\frac{b}{r}\right)\left(2\Phi''+2\Phi'^2\right)+\frac{\Phi'^2}{r^2}\left(b'r+3b-4r\right)+\frac{2b'}{r^2}.
\end{equation}
This will equal to
\begin{equation}
8\pi T=8\pi\left[(\rho-3p)-\frac{3}{2\lambda}\left(\rho^2+3p^2-\frac{1}{3}(\rho-3p)^2\right)\right]
\end{equation}
In an orthonormal reference frame the components of the projected Weyl tensor  $\varepsilon _{\hat{\mu }{\hat{\nu}}}$ have the components
\begin{equation}
\varepsilon _{\hat{\mu }{\hat{\nu}}}={\rm diag}\left[\epsilon (r), \sigma _r (r), \sigma _t (r),\sigma _t (r)\right].
\end{equation}
The components of the effective energy-momentum tensor have the form $\rho ^{{\rm eff}}=\rho \left(1+\rho /2\lambda \right)-\epsilon /8\pi$, $p ^{{\rm eff}}_r=p \left(1+\rho /\lambda \right)+\rho ^2/2\lambda-\sigma _r /8\pi$ and $p ^{{\rm eff}}_t=p \left(1+\rho /\lambda \right)+\rho ^2/2\lambda-\sigma _t /8\pi$, respectively. The NEC violation, $\rho ^{{\rm eff}}+p ^{{\rm eff}}_r<0$ provides the following generic restriction,
\begin{equation}
8\pi \left(\rho +p\right)\left(1+\frac{\rho }{\lambda }\right)<\epsilon +\sigma _r.
\end{equation}
Hence braneworld gravity provides a natural scenario for the existence of traversable wormholes \cite{Lobo}.

\section{4D study of the inflating braneworld wormhole model}\label{infbwh}

Roman \cite{Roman} has suggested that inflation might provide a natural mechanism for the enlargement of Planck size wormholes to a macroscopic size.
In the following we consider the inflation of the wormholes in the framework of the brane world models.

\subsection{Brane geometry and the energy-momentum tensor}

The metric of an inflating wormhole on the brane is given by
\begin{equation}
ds^2=-e^{2\Phi(r,t)}dt^2+a^2(t)\left[\frac{dr^2}{1-b(r)/r}+r^2(d\theta^2+\sin^2\theta d^2\phi)\right].\label{wormholemetric}
\end{equation}
In Eq.~(\ref{wormholemetric}) we have introduced a dynamical redshift function, but we do not consider the form function to evolve with time. This is because a nonzero $db(r_0)/dt$ would require an infinite energy density of the matter at the throat.  With this metric, the components of the Einstein tensor are given by
\begin{equation}
G_{tt}=3\left(\frac{\dot{a}}{a}\right)^2+\frac{e^{2\Phi}}{a^2}\frac{b'}{r^2},\label{Gtt}
\end{equation}
\begin{equation}
G_{tr}=\frac{2\dot{a}\Phi'}{a},
\end{equation}
\begin{equation}
G_{rr}=\frac{r}{r-b}\Bigl[2\dot{\Phi}a\dot{a}e^{-2\Phi}-2a\ddot{a}e^{-2\Phi}-\dot{a}^2e^{-2\Phi}+ \frac{2\Phi'}{r}\bigl(1-\frac{b}{r}\bigr)-\frac{b}{r^3}\Bigr],
\end{equation}
\begin{eqnarray}
G_{\theta\theta}&=&r^2\left[e^{-2\Phi}\bigl[2a\dot{a}\dot{\Phi}-2a\ddot{a}-\dot{a}^2\bigr]+ (\Phi''+\Phi'^2)\left(1-\frac{b}{r}\right)\right.+\nonumber\\
&&\left.\frac{\Phi'}{r}\left(1-\frac{b}{2r}-\frac{b'}{2}\right)+ \frac{1}{2r^2}\left(\frac{b}{r}-b'\right)\right],
\end{eqnarray}
\begin{eqnarray}
G_{\phi\phi}&=&r^2\sin^2\theta \left[e^{-2\Phi}\bigl[2a\dot{a}\dot{\Phi}-2a\ddot{a}-\dot{a}^2\bigr]+ (\Phi''+\Phi'^2)\left(1-\frac{b}{r}\right)\right.+\nonumber\\
&&\left.\frac{\Phi'}{r}\left(1-\frac{b}{2r}-\frac{b'}{2}\right)+ \frac{1}{2r^2}\left(\frac{b}{r}-b'\right)\right].\label{Gpp}
\end{eqnarray}

We assume that the inflation is driven by a homogenous scalar field $\phi $,  filling the Universe, and with energy- momentum tensor given by Eq.~(\ref{scalarfieldT}). The total matter contribution $T^{\rm{mat}}_{\mu\nu}=T_{\mu\nu}+(6/\lambda) S_{\mu\nu}$ on the brane to the energy-momentum tensor in Eq.~(\ref{einstein}) is given by
\begin{eqnarray}
T^{\rm{mat}}_{tt}&=&e^{2\Phi}\left(e^{-2\Phi}\frac{\dot{\phi}^2}{2}+V+e^{-4\Phi}\frac{\dot{\phi}^4}{8\lambda}+ e^{-2\Phi}\frac{\dot{\phi}^2V}{2\lambda}+\frac{V^2}{2\lambda}\right),\label{tmattt}\\
T^{\rm{mat}}_{rr}&=&\frac{a^2r}{r-b}\left(e^{-2\Phi}\frac{\dot{\phi}^2}{2}-V+ e^{-4\Phi}\frac{3\dot{\phi}^4}{8\lambda}+e^{-2\Phi}\frac{\dot{\phi}V}{2\lambda}-\frac{V^2}{2\lambda}\right),\\
T^{\rm{mat}}_{\theta\theta}&=&a^2r^2\left(e^{-2\Phi}\frac{\dot{\phi}^2}{2}-V+ e^{-4\Phi}\frac{3\dot{\phi}^4}{8\lambda}+e^{-2\Phi}\frac{\dot{\phi}V}{2\lambda}-\frac{V^2}{2\lambda}\right),\\
T^{\rm{mat}}_{\phi\phi}&=&a^2r^2\sin^2\theta\left(e^{-2\Phi}\frac{\dot{\phi}^2}{2}-V+ e^{-4\Phi}\frac{3\dot{\phi}^4}{8\lambda}+e^{-2\Phi}\frac{\dot{\phi}V}{2\lambda}- \frac{V^2}{2\lambda}\right)\label{tmatpp}.
\end{eqnarray}

\subsection{The nonlocal projected Weyl tensor}

 The components of the projection of the bulk Weyl tensor $C_{ABCD}$ on the brane can be obtained from Eqs.~ (\ref{Gtt}) - (\ref{Gpp}) and Eqs.~(\ref{tmattt}) - (\ref{tmatpp}), respectively. These equations give the explicit form of $\varepsilon_{\mu\nu}$ in terms of the wormhole's redshift and form functions, and of the scalar field, respectively. The $tt$, $tr$ and $rr$ components of $\varepsilon_{\mu\nu}$ are given by
\begin{equation}\label{ett}
\varepsilon_{tt}=8\pi e^{2\Phi}\left(e^{-2\Phi}\frac{\dot{\phi}^2}{2}+V+e^{-4\Phi}\frac{\dot{\phi}^4}{8\lambda}+e^{-2\Phi}\frac{\dot{\phi}^2V}{2\lambda}+\frac{V^2}{2\lambda}\right)-\left[3\left(\frac{\dot{a}}{a}\right)^2+\frac{e^{2\Phi}}{a^2}\frac{b'}{r^2}\right],
\end{equation}
\begin{equation}
\varepsilon_{tr}=-\frac{2\dot{a}\Phi'}{a},
\end{equation}
\begin{eqnarray}\label{err}
\varepsilon_{rr}&=&8\pi\frac{a^2r}{r-b}\left(e^{-2\Phi}\frac{\dot{\phi}^2}{2}-V+ e^{-4\Phi}\frac{3\dot{\phi}^4}{8\lambda}+e^{-2\Phi}\frac{\dot{\phi}V}{2\lambda}-\frac{V^2}{2\lambda}\right)-\nonumber\\
&&\frac{r}{r-b}\left[2\dot{\Phi}a\dot{a}e^{-2\Phi}-2a\ddot{a}e^{-2\Phi}- \dot{a}^2e^{-2\Phi}+\frac{2\Phi'}{r}\left(1-\frac{b}{r}\right)-\frac{b}{r^3}\right].
\end{eqnarray}
Taking the trace of the brane Einstein equation Eq.~(\ref{einstein}) gives another constraint equation for the Ricci tensor,
\begin{eqnarray}\label{con2}
&&6e^{-2\Phi}\left[\frac{\ddot{a}}{a}-\dot{\Phi}\frac{\dot{a}}{a}+\left(\frac{\dot{a}}{a}\right)^2\right]+8\pi\left[e^{-4\Phi}\frac{\dot{\phi}^4}{\lambda}+e^{-2\Phi}\frac{\dot{\phi}^2V}{\lambda}+e^{-2\Phi}\dot{\phi}^2-4V-\frac{2V^2}{\lambda}\right]-\nonumber\\
&&\left(1-\frac{b}{r}\right)\left(\frac{2\Phi''}{a^2}+\frac{2\Phi'^2}{a^2}\right)+\frac{\Phi'^2}{r^2a^2}\left(b'r+3b-4r\right)+\frac{2b'}{r^2a^2}=0.
\end{eqnarray}

\subsection{The energy conditions}

Similarly to the case of the static redshift inflating wormhole with $\Phi(r,t)=\Phi(r)$, in order for the wormhole to be asymptotically flat the "flaring out condition" must be satisfied. The metric of constant time and $\theta=\pi /2$ slice in the space defined by Eq.~(\ref{wormholemetric}) is
\begin{equation}\label{h1}
ds^3=\frac{a^2(t)dr^2}{1-b(r)/r}+a^2(t)r^2d\Omega^2.
\end{equation}
With the use of Eq.~(\ref{h1}) we can formulate the flaring out condition as \cite{Roman}
\begin{equation}
\frac{b-b'r}{2b^2}>0.
\end{equation}
The energy condition  at the throat can be worked out by considering the scalar quantity $\lim_{r\rightarrow r_0}T^{\rm eff}_{\mu\nu}W^{\mu}W^{\nu}$, where $T^{\rm eff}_{\mu\nu}=T_{\mu\nu}+\frac{6}{\lambda}S_{\mu\nu}-\frac{\varepsilon_{\mu\nu}}{8\pi}$ is the effective energy momentum tensor on the brane. For a radial outgoing null vector $W^{\mu}=(e^{-\Phi},\pm \frac{\sqrt{1-b/r}}{a},0,0)$ we obtain
\begin{equation}\label{energy}
\lim_{r\rightarrow r_0}T^{\rm eff}_{\mu\nu}W^{\mu}W^{\nu}=2\left[\left(\frac{\dot{a}}{a}\right)^2-\frac{\ddot{a}}{a}+\dot{\Phi}\frac{\dot{a}}{a}\right]e^{-2\Phi}+\frac{b'-1}{a^2r_0^2}.
\end{equation}
Therefore the energy conditions also evolve with the cosmological expansion of the Universe.

\section{Brane evolution of the inflating wormhole}\label{4}

The evolution of the scale factor $a(t)$ and of the scalar field $\phi(t)$ are governed by the global evolution of the Universe. For an expanding wormhole the asymptotic behaviors of the form function and of the redshift function are given by $\lim_{r\rightarrow\infty}b(r)/r=0$ and $\lim_{r\rightarrow\infty}\Phi(r)=0$, respectively.  We also impose the condition  $\lim_{r\rightarrow \infty}b'/r^2=0$, so that asymptotically the behavior of the metric of the wormhole is the same as the behavior of the metric of the global braneworld model. Taking these limits in the dynamical equations describing the scale factor and the scalar field evolution gives the basic equations describing the inflating wormhole on the brane as
\begin{equation}\label{con3}
3\left(\frac{\dot{a}}{a}\right)^2=8\pi\left[\frac{\dot{\phi}^2}{2}+V+ \frac{\dot{\phi}^4}{8\lambda}+\frac{\dot{\phi}^2V}{2\lambda}+\frac{V^2}{2\lambda}\right],
\end{equation}
and
\begin{equation}\label{con4}
\ddot{\phi}+3\frac{\dot{a}}{a}\dot{\phi}+V'(\phi)=0,
\end{equation}
respectively. Since the two variables in these equations $a(t)$ and $\phi(t)$ are functions of $t$ only, we expect that these two equations actually hold everywhere in the wormhole.

The dynamics of a scalar field in the brane world cosmology were studied in \cite{Mizuno} and \cite{Maartens}, respectively, by assuming that the scalar field is confined in the 4-dimensional world. As for the potential of the scalar field, several types of potentials were considered.  It has been shown that when the energy density square term is the dominating term in the energy-momentum tensor, the behavior of the scalar field is very different from the conventional cosmology. In the following we will restrict the discussion of the inflating wormhole to the scalar field potential
\begin{equation}
V(\phi)=\frac{1}{2}m^2\phi^2,
\end{equation}
which corresponds to the chaotic inflation model on the brane\cite{Maartens, Mizuno}.

\subsection{Time evolution of the redshift function}\label{timeevolve}

The dynamic of the redshift function of a braneworld wormhole is given by Eq.~(\ref{bulkcon3}). Using the expressions of the metric tensor components on the induced metric we obtain
\begin{equation}\label{phidot}
\dot{\Phi}\dot{\phi}=V'(\phi)\left(e^{2\Phi}-1\right).
\end{equation}
For $e^{2\Phi}\ne 1$, we can separate the variables of the above equation into
\begin{equation}
\frac{\dot{\Phi}}{e^{2\Phi}-1}=\frac{V'(\phi)}{\dot{\phi}}.
\end{equation}
With
\begin{equation}
\int{\frac{dx}{e^{2x}-1}}=\ln{\sqrt{\frac{e^{2x}-1}{e^{2x}}}},
\end{equation}
and by using $V(\phi)=m^2\phi^2/2$, we obtain for the redshift function the expression
\begin{equation}
e^{-2\Phi(r,t)}=1-e^{\int^t_{t_0}{\frac{2m^2\phi dt}{\dot{\phi}}}}\left[1-e^{-2\Phi(r,t_0)}\right],\label{evoPhi}
\end{equation}
where $\Phi(r,t_0)$ is the initial redshift function, which must be finite everywhere. The asymptotic behavior of the redshift function is given by  $\lim_{r\rightarrow\infty}\Phi(r,t)=1$, and this behavior is the same for all times. The time evolution of the redshift function at the throat is described by the equation
\begin{equation}\label{evoP}
e^{2\Phi(r_0,t)}=\frac{1}{1-A(r_0)e^{\int^t_{t_0}{\frac{2m^2\phi dt}{\dot{\phi}}}}},
\end{equation}
where $A(r)=\left[1-e^{-2\Phi(r,t_0)}\right]$ can be obtained from the initial conditions. By taking the limit of small $A$ we obtain
\begin{equation}
e^{2\Phi(r_0,t_0)}=\frac{1}{1-A(r_0)}\approx 1+A(r_0)\Rightarrow \psi_{\rm Newton}=A(r_0).
\end{equation}
This equation shows that $A(r)$ is actually the Newtonian potential, and therefore we assume that $A(r)\leq 0$.

\section{A simple solution - the zero redshift case}\label{zeroredshift}

If the initial condition of the wormhole is so that $\Phi(r, t_0)=0$, Eq.~(\ref{phidot}) implies that $\Phi=0$ for all times. Then we also obtain
\begin{equation}
\frac{b'}{r^2a^2}=0.
\end{equation}
Since $b(r_0)=r_0$, the solution for the form function is  $b(r)=r_0$. For this case the non - zero components of the Einstein tensor are given by
\begin{eqnarray}
G_{tt}&=&3\left(\frac{\dot{a}}{a}\right)^2,\label{Gtt1}\\
G_{rr}&=&\frac{a^2r}{r-r_0}\left[-2\frac{\ddot{a}}{a}-\frac{\dot{a}^2}{a^2}-\frac{r_0}{a^2r^3}\right],\\
G_{\theta\theta}&=&a^2r^2\left[-2\frac{\ddot{a}}{a}-\frac{\dot{a}^2}{a^2}+\frac{r_0}{2a^2r^3}\right],\\
G_{\phi\phi}&=&a^2r^2\sin^2(\theta)\left[-2\frac{\ddot{a}}{a}- \frac{\dot{a}^2}{a^2}+\frac{r_0}{2a^2r^3}\right].\label{Gpp1}
\end{eqnarray}

The energy - momentum tensor of the scalar field on the brane world wormhole is obtained as
\begin{eqnarray}
T_{\mu\nu}&=&{\rm diag}\left\{\frac{1}{2}\dot{\phi}^2+V(\phi), \frac{a^2r}{r-r_0}\left[\frac{1}{2}\dot{\phi}^2-V(\phi)\right],\right.\\
&&\left.a^2r^2\left[\frac{1}{2}\dot{\phi}^2-V(\phi)\right], a^2r^2\sin^2(\theta)\left[\frac{1}{2}\dot{\phi}^2-V(\phi)\right]\right\}.
\end{eqnarray}

 Eqs.~(\ref{ett}) and Eq.~(\ref{err}) give the exact form of the projected Weyl tensor on the brane as
\begin{equation}\label{e1}
\varepsilon_{\mu\nu}={\rm diag}\left(0, -\frac{r_0}{r^3\left(1-\frac{r_0}{r}\right)}, \frac{r_0}{2r}, \frac{r_0}{2r}\right).
\end{equation}

One can easily verify that $g^{\mu\nu}\varepsilon_{\mu\nu}=0$. In the local coordinate system, the projected Weyl tensor can be represented as $\varepsilon _{\mu \nu}={\rm diag}\left(\epsilon_0, \epsilon_r, \epsilon_t, \epsilon_t\right)$, with
\begin{equation}
\epsilon_0=0, \epsilon_r=-\frac{r_0}{r^3a^2}, \epsilon_t=\frac{r_0}{2r^3a^2}.
\end{equation}
We concluded that the Weyl contribution that supporting the wormhole decay as $a^{-2}$.
\section{Chaotic inflation in the braneworld model}\label{6}

According to \cite{Maartens, Mizuno}, in the case of the chaotic inflation driven by the scalar field potential $V(\phi)=m^2\phi^2/2$, the expressions of $a(t)$ and $\phi(t)$ can be approximated by two different sets of expressions, corresponding to the exponential inflationary period,  and to the final stages of inflation, respectively.

\subsection{The Universe during exponential inflation}\label{dinf}

Let $t_i$ be the initial time of inflation, and let $t_f$ to be the end time of the inflation. During inflation, the potential term dominates the kinetic energy term, so that $V(\phi)>>\dot{\phi}^2$, and $V'>>\ddot{\phi}$, respectively. Then Eqs.~(\ref{con3}) and (\ref{con4}) have the solutions
\begin{eqnarray}
\phi (t)&=&\left(\frac{4\lambda}{3\pi}\right)^{1/4}(t_1-t)^{1/2},\label{midinf1}\\
a(t)&=&a_1 \exp\left[-\frac{m^2}{3}(t_1-t)^2\right],\label{midinf2}
\end{eqnarray}
where $t_1$ and $a_1$ are arbitrary integration constants, which can be obtained from the initial conditions, and $m$ is the 5D Planck mass, respectively. The end of inflation is determined by the breaking down of the slow roll condition, $V(\phi)\approx\dot{\phi}^2$, or $V'\approx\ddot{\phi}$. These two conditions fix the value of $t_f$, which is given by the expression $t_f\approx t_1-(2m)^{-1}$. By using this expression for $t_f$, we obtain the scale factor after the exponential inflation phase as
\begin{equation}
a_f:=a\left(t_f\right)=a_1\exp\left[-\frac{m^2}{3}(2m)^{-2}\right]=a_1\exp\left(-\frac{1}{12}\right).
\end{equation}
Since $t_i<<t_f$, if we adopt the e-folding number for the exponential inflation phase to be $60$, we obtain the following estimate for the ratio of the scale factors at the beginning and end of inflation, respectively,
\begin{equation}
\frac{a\left(t_f\right)}{a\left(t_i\right)}=\exp\left[-\frac{1}{12}+\frac{m^2t_1^2}{3}\right]\approx \exp(60)
\end{equation}
Therefore $mt_1\approx \sqrt{180.25}\approx 13.43$, and $mt_f\approx 12.9$.

\subsection{The oscillating scalar field phase}

 After the end of the exponential inflation phase, the scalar field enters in an oscillating stage, and it keeps oscillating even after the radiation domination era. The oscillating solution is approximated by \cite{Maartens, Mizuno}
\begin{eqnarray}
\phi&=&\left(\frac{\lambda}{3\pi}\right)^{\frac{1}{4}}\frac{\sin(mt)}{mt^{1/2}},\label{end1}\\
a&=&a_2 (t/t_2)^{1/3}\label{end2},
\end{eqnarray}
where $t_2$ and $a_2$ are arbitrary integration constants. We can choose $t_2$ to be the time for which the oscillating stage begins, and determine it from the initial conditions. For example, if at the beginning of the oscillating stage the scalar field is one fourth of that at the end of the exponential inflation, then
\begin{equation}\label{interp}
\frac{1}{4}\left(\frac{4\lambda}{3\pi}\right)^{\frac{1}{4}}\left(t_1-t_f\right)^{\frac{1}{2}}= \left(\frac{\lambda}{3\pi}\right)^{\frac{1}{4}}\frac{\sin\left(mt_2\right)}{mt_2^{1/2}}.
\end{equation}
Together with $t_f\approx t_1-(2m)^{-1}$, this equation gives the following algebraic equation for $mt_2$,
\begin{equation}
\frac{mt_2}{16}-\sin^2(mt_2)=0
\end{equation}
By using Newton's method, we find that the zero of this equation next to $t_f$ is given by $mt_2\approx 13.75$. We will discuss later that the solution is insensitive
to the way we choose to interpret the two stages.

\section{The evolution of the inflating wormhole}\label{soln}

In the present Section we consider the evolution of the inflating brane world wormhole during the different phases of inflation

\subsection{Wormhole evolution during the exponential inflation phase}\label{7}

With the use of  Eqs.~(\ref{midinf1}) and Eq.~(\ref{midinf2}), the integral in Eq.~(\ref{evoP}) becomes
\begin{equation}\label{int1}
\int^t_{t_i}{\frac{2m^2\phi dt}{\dot{\phi}}}=\int^t_{t=t_i}m^2(t-t_1)dt=\int^{x=mt}_{x=mt_i}{(x-mt_1)dx}.
\end{equation}
Thus, for example, a $60$ e-folding exponential inflation gives the following solution, describing the time evolution of the redshift function of the inflating brane world wormhole
\begin{equation}\label{redshiftmid}
e^{2\Phi(r,t)}=\frac{1}{1-A(r)\exp\left(\frac{m^2t^2}{2}-m^2t_1t\right)},\; {\rm for}\; mt<mt_f=12.9.
\end{equation}
The time evolution of the energy condition at the throat during the inflationary phase can then be obtained from Eq.~(\ref{energy}) and from the global evolution Eq.~(\ref{midinf1}) as
\begin{eqnarray}\label{energy_inf}
\lim_{r\rightarrow r_0}T^{\rm eff}_{\mu\nu}W^{\mu}W^{\nu}&=&\frac{4m^2}{3}\left[1-A(r_0)e^{\frac{m^2t^2}{2}-m^2t_1t}\right]-\frac{2m^2}{3}A(r_0)(mt-mt_1)^2e^{\frac{m^2t^2}{2}-m^2t_1t}+\nonumber\\
&&\frac{b'-1}{a_1^2r_0^2}e^{\frac{2}{3}(mt-mt_1)^2},\mbox{ for }mt \leq 12.9.
\end{eqnarray}
In this phase, we see that the among of exotic matter is diluting as an exponential function of the time. Furthermore, there is a range of wormhole initial conditions such that the wormhole could survive the exponential inflationary phase, that is, the energy condition on the effective matter remain after the exponential inflation. At the end of inflation, the energy condition is obtained by substituting $mt=12.9$ into Eq.~(\ref{energy_inf}),
\begin{equation}
\lim_{r\rightarrow r_0}T^{\rm eff}_{\mu\nu}W^{\mu}W^{\nu}|_{t=t_f}\approx\frac{4m^2}{3}\left[1-9.3\times10^{-40}A(r_0)\right]+\frac{b'-1}{a^2(t_f)r_0^2}<0,
\end{equation}
where we have also used the value of $mt_1$ obtained earlier. For example, wormholes with shaping function constrained by the following equation could survive the exponential inflation,
\begin{equation}
b'<-\frac{4m^2a^2(t_f)r_0^2}{3}+1.
\end{equation}

\subsection{Dynamical evolution of the wormhole in the oscillating scalar field phase}

To obtain the evolution of the wormhole near the end of the exponentially inflation phase, corresponding to $t>t_2$, we substitute Eq.~(\ref{end1}) and Eq.~(\ref{end2}) into the global evolution equation Eq.~(\ref{evoP}). The integral in Eq.~(\ref{evoP}) becomes
\begin{equation}\label{integral}
\int^t_{t_2}{\frac{2m^2\phi dt}{\dot{\phi}}}=\int_{x=mt_2}^{x=mt}{\frac{4x\tan xdx}{2x-\tan x}}.
\end{equation}
Recall that $t_2$ is the time at which the oscillating stage begins. We can see  that there is a singularity in the integral, no matter when we switch to the oscillating stage. The evolving wormhole will eventually face a collapsing phase. To numerically analyze the solution, we adopt the $60$ e-folding and the initial condition given by Eq.~(\ref{interp}), and we assume that the value of the redshift function in Eq.~(\ref{redshiftmid}) at the end of the exponential inflation is the initial value for this phase. Hence the evolution of the redshift function at the throat can be written as
\begin{equation}\label{integral2}
e^{2\Phi(r_0,t)}=\frac{1}{1-A(r_0)e^{-90}\exp{\int_{x=13.75}^{x=mt}{4x \tan xdx/\left(2x-\tan x\right)}}},\;{\rm for}\; x<x_c,
\end{equation}
where $x_c$ is the first singular point of the function $4x\tan x/\left[2x-\tan x\right]$ that is larger than $13.75$, i.e., the first root of the equation $2x-\tan(x)=0$ greater than $13.75$. With Newton's method we obtain that $x_c\approx 14.101725$, which is e-folding dependent. Besides, the value $e^{-90}$ is obtained from the exponent $\exp{\left(m^2t^2/2-13.43 mt\right)}$ in Eq.~(\ref{redshiftmid}) at $mt=12.9$.

Based on the singular behavior at $x_c$ of the integral in Eq.~(\ref{integral}),  we can derive the fate of a wormhole
at the characteristic time $x_c$ for different initial conditions $A(r_0)$.


\paragraph{$A\left(r_0\right)=0$.}

This initial condition describes the  zero redshift wormhole. It is the only initial condition that makes $\Phi$ to remain finite everywhere. However, due to the fact that the integral $\int_{x=13.75}^{x=x_c}{4x \tan xdx/\left(2x-\tan x\right)}$ is unbounded, it follows that such an initial condition is very restrictive. It is because any small derivation from zero redshift would be enlarged. On the other hand, the energy condition for the effective matter on the brane is violated at the throat for all times. It is the Weyl contribution from the bulk supporting such violation. As we have seen in Section~\ref{zeroredshift}, this Weyl contribution decay as $a^{-2}$.

\paragraph{$A\left(r_0\right)<0$.}

This initial condition makes the redshift function $e^{2\Phi}$ vanish, since the integral in Eq.~(\ref{integral2}) diverges.
This case corresponds to the appearance of an event horizon at the throat of the wormhole, and it can be interpreted as a conversion of wormhole into a black hole \cite{Ku08}. Once the wormhole is converted into a black hole, the metric Eq.~(\ref{wormholemetric}) does not describe anymore the geometry. This result also shows that any dynamical wormhole that survives the exponential inflation phase would collapse to a black hole, immediately after the oscillating stage begins. To see how the energy condition evolves in this stage, we analyze the $\dot{\Phi}$ terms in Eq.~(\ref{energy}). Since $e^{-2\Phi}\rightarrow\infty$, there must be one moment in which the $2\dot{\Phi}e^{-2\Phi}$ term diverges, and it is negative as $\Phi$ decreases. Therefore the energy condition of the effective matter remain violated during the conversion, which may form a unique signature on this black hole.

\section{Discussions and final remarks}\label{astroimp}

In the present paper we have considered the evolution of inflating wormholes in the braneworld model. As a first step in our study we have considered the conditions under which a braneworld wormhole can be embedded into a five dimensional bulk. We have found that such an embedding is, at least in principle, always possible. As a general result, we have found that braneworld wormholes created in the early Universe  would experience a transition to a black hole. If there is a large number of braneworld wormholes created in the early Universe, that would lead to the creation of a large number of black holes, and  this would have significant astrophysical implications for the global evolution of the Universe, including the large scale structure formation period.

Soon after the big bang, due to the symmetry breaking, the Universe underwent a phase transition after which it was filled up with a highly homogenous scalar field. However, space-time inhomogeneities can still be generated in such a homogenous background. For example, the perturbations of the scalar field could generate the initial inhomogeneities that eventually would be responsible for structure formation. Another possibility for generating some very early space-time inhomogeneities during inflation would be the presence of braneworld wormholes in the early Universe.  The braneworld wormholes are solutions of the Einstein equations on the brane. In the braneworld scenario, the existence of these wormholes does not necessarily leads to the violation of the weak energy conditions. Braneworld  wormholes would evolve, and increase in size as the Universe expands. With the exception of the particular case in which the wormholes have exactly zero redshift everywhere,  when the Universe switches from the exponential inflation to the oscillating stage, a very small initial value $A\left(r_0\right)$ of the redshift function would make the wormhole to collapse into a black hole. The transition wormhole - black hole is realized through the sudden appearance of an event horizon. Since these black holes formed at a very early stage in the evolution of the Universe, and since there are no lower limits on their size, the black holes created from a wormhole should be considered as primordial. Once a wormhole becomes a black hole, it will soon lose its co-expansion with the Universe, and it can gain mass through relatively slow accretion during the radiation dominated era of the Universe \cite{Carr1, Carr2}.

Suppose that the Universe was filled with wormholes that are separated  by a distance $r_e$. We would like to estimate the throat radius, so that these wormholes do not significantly change the energy content of the universe. In order for this condition to be satisfied, we compare the total energy of a wormhole with the energy of the cosmological background. From Eq.~(\ref{Gtt}) we know that if we consider the average over whole space, the effect of creating a wormhole is similar to adding a extra energy term to the Universe that decays as $1/a^2r^2$. If we impose the condition that this extra energy, necessary to create the wormhole, is larger than the energy corresponding to a wormhole filled with the cosmological background energy, we obtain the condition
\begin{equation}
\int^{r_e}_{r_0}{\frac{e^{2\Phi}b'}{a^2r^2}a^3d^3r}>\int^{r_0}_{0}{\rho a^3d^3r},
\end{equation}
where $\rho$ is the energy of the cosmological background. To obtain an estimate of the wormhole throat after inflation we use the form function of the wormhole for the dust solution,  $b'=\gamma r_0^2/r^2$, which gives
\begin{equation}
4\pi\frac{\gamma^2 r_0}{a^2}> \rho\frac{4}{3}\pi r_0^3.
\end{equation}
Since $0<\gamma<1$, in this order of magnitude estimation we take $\gamma \approx 1$. If we adopt a reheating temperature of $T_{\rm RH}=10^9\;{\rm GeV}^4$ \cite{Cyburt,Ellis}, we obtain for the density of the cosmological background the approximate value  $\rho\approx0.6\times10^{72}\;{\rm eV}^4$. By choosing the end of inflation at a value of the scale factor of the order of $a\approx10^{-20}$  \cite{Wong}, we can obtain an estimate of $r_0$ as
\begin{equation}
r_0<\sqrt{\frac{3}{\rho\pi a^2}}\approx10^7\;{\rm cm}.
\end{equation}
The mass limit corresponding to a Schwarzschild's black hole would be
\begin{equation}
M_0<\frac{10^2\;{\rm km}}{3\;{\rm km}}M_{\odot}\approx6.6\times10^{34}\;{\rm g},
\end{equation}
corresponding to around $30$ solar masses.

In the present paper we have also explored some specific physical properties and characteristics of the inflating wormhole geometries, and we have found the
expressions describing the properties of the wormholes during the different phases of the inflation. In a
future work we will consider in more details the possible astrophysical implications of our results.

\section{Acknowledgments}

We would like to thank to the anonymous referees for comments and suggestions that helped us to significantly improve the manuscript. The authors would like to thank Dr. Francisco S. N. Lobo for useful suggestions and discussions. The work described in this paper was supported by a grant from the Research Grants Council of the Hong Kong Special Administrative Region, China (Project No. HKU 701808P).


\begin{thebibliography}{99}

\bibitem{Morris:1988cz}
  M.~S.~Morris and K.~S.~Thorne,
  Am.\ J.\ Phys.\  {\bf 56}, 395 (1988).

\bibitem{Visser}
  M.~Visser,
  Lorentzian wormholes: from Einstein to Hawking,
  AIP Press, (1995).

\bibitem{Morris2} M. S. Morris, K. S. Thorne, and U. Yurtsever, Phys. Rev. Lett. \textbf{61}, 1446 (1988)

\bibitem{Frolov} V. P. Frolov and I. D. Novikov, Phys. Rev. \textbf{D42}, 1057, (1990)

\bibitem{solutions}
 J.~P.~S.~Lemos, F.~S.~N.~Lobo and S.~Quinet de Oliveira,
  Phys.\ Rev.\ {\bf D68}, 064004 (2003).
%
  F.~S.~N.~Lobo,
  Phys.\ Rev.\ {\bf D71}, 084011 (2005);
  %
  F.~S.~N.~Lobo,
  Phys.\ Rev.\ {\bf D71}, 124022 (2005);
P. K. F. Kuhfittig, Phys. Rev. {\bf D73},  084014 (2006);
 F.~S.~N.~Lobo,
  Phys.\ Rev.\ {\bf D75}, 064027 (2007);
%
F.~S.~N.~Lobo,
  Phys.\ Rev.\ {\bf D73}, 064028 (2006);
C. G. Boehmer, T. Harko, F. S. N. Lobo, Class. Quant. Grav. {\bf 25}, 075016 (2008); F. S. N. Lobo, Class. Quant. Grav. 25, 175006 (2008); T. Harko, Z. Kovacs, F. S. N. Lobo, Phys. Rev. {\bf D78}, 084005 (2008); R. Garattini and  F. S. N. Lobo, Phys. Lett. {\bf B671}, 146 (2009); T. Harko, Z. Kovacs, F. S. N. Lobo, Phys. Rev. {\bf D79}, 064001 (2009); F. S. N. Lobo and M. A. Oliveira, Phys. Rev. {\bf D80}, 104012 (2009); N. Montelongo Garcia and F. S. N. Lobo, Phys. Rev. {\bf D82}, 104018 (2010); M. Jamil, P. K. F. Kuhfittig, F. Rahaman, and S. A. Rakib,  Eur. Phys. J. C {\bf 67},  513 (2010).

\bibitem{Morris:1988tu}
  M.~S.~Morris, K.~S.~Thorne and U.~Yurtsever,
  Phys.\ Rev.\ Lett.\  {\bf 61}, 1446 (1988).

\bibitem{Hawking} S. W. Hawking and G. F. R. Ellis, The Large Scale Structure of Spacetime (Cambridge Univeristy Press, London, 1973), pp.88-96.

\bibitem{Tipler} F. J. Tipler, Phys. Rev. \textbf{D17}, 2521 (1978); T. Roman, ibid. \textbf{33}, 3526 (1986); A. Borde, Class. Quantum Grav. \textbf{4}, 343 (1987)


\bibitem{Boehmer:2007rm}
  C.~G.~Boehmer, T.~Harko and F.~S.~N.~Lobo,
  Phys. Rev. {\bf D76} 084014 (2007).
\bibitem{Roman} T. A. Roman, Phys. Rev. \textbf{D47}, 1370 (1993).

\bibitem{Randall}  L. Randall and R. Sundrum, Phys. Rev. Lett \textbf{83},
3370 (1999).

\bibitem{Randall2}  L. Randall and R. Sundrum, Phys. Rev. Lett \textbf{83},
4690 (1999).

\bibitem{Shiromizu}  M. Sasaki, T. Shiromizu and K. Maeda, Phys. Rev.
\textbf{D62}, 024008 (2000); T. Shiromizu, K. Maeda and M. Sasaki, Phys.
Rev. \textbf{D62}, 024012 (2000); K. Maeda, S. Mizuno and T. Torii, Phys.
Rev. \textbf{D68}, 024033 (2003).

\bibitem{all2}
 A. Campos and C. F. Sopuerta, Phys. Rev. {\bf D63}, 104012
(2001);  C.-M. Chen, T. Harko and M. K. Mak, Phys. Rev. {\bf D64},
044013 (2001);  C.-M. Chen, T. Harko and M. K. Mak, Phys. Rev. {\bf D64},
124017 (2001); J. D. Barrow and R. Maartens, Phys. Lett. {\bf
B532}, 153 (2002); C.-M. Chen, T. Harko, W. F. Kao and M. K. Mak,
Nucl. Phys. {\bf B64}, 159 (2002);  T. Harko and
M. K. Mak, Class. Quantum Grav. {\bf 20}, 407 (2003); C.-M. Chen,
T. Harko, W. F. Kao and M. K. Mak, JCAP {\bf 0311}, 005 (2003); T.
Harko and M. K. Mak, Class. Quantum Grav. {\bf 21}, 1489 (2004);
M. K. Mak and T. Harko, \prd {\bf 70}, 024010 (2004); T. Harko and
M. K. Mak, Phys. Rev. {\bf D69}, 064020 (2004); A. N. Aliev and A.
E. Gumrukcuoglu, Class. Quant. Grav. {\bf 21}, 5081 (2004); M.
Maziashvili, Phys. Lett. {\bf B627}, 197 (2005);  M. K. Mak and T. Harko, Phys.
Rev. {\bf D71}, 104022 (2005);  T. Harko and K. S. Cheng,
Astrophys. J. {\bf 636}, 8 (2006); L. A. Gergely, Phys. Rev. {\bf
D74} 024002, (2006); C. G. B\"ohmer and T. Harko, Class.
Quantum Grav. {\bf 24}, 3191 (2007); M. Heydari-Fard and H. R.
Sepangi, Phys. Lett. {\bf B649}, 1 (2007); T. Harko and K. S.
Cheng, Phys. Rev. {\bf D76}, 044013 (2007); A. Viznyuk and Y.
Shtanov, Phys. Rev. {\bf D76}, 064009 (2007); Z. Kovacs and L. A.
Gergely, Phys. Rev. {\bf D77}, 024003 (2008); T. Harko and V. S.
Sabau, Phys. Rev. {\bf D77}, 104009 (2008); Z. Keresztes and L. A. Gergely, Ann. Physik {\bf 19}, 249 (2010); Z. Keresztes and L. A. Gergely,
Class. Quant. Grav. {\bf 27}, 105009 (2010); L. A. Gergely, T. Harko, M. Dwornik, G. Kupi, and Z. Keresztes, to appear in MNRAS, arXiv:1105.0159 (2011).

\bibitem{Umezu} K. Umezu, K. Ichiki, T. kajino, G. J. Mathews, R. Nakamura, and M. Yahiro, Phys. Rev. \textbf{D73}, 063527 (2005).

\bibitem{Ap}  P. S. Apostolopoulos and N. Tetradis, Phys. Rev. \textbf{D71},
043506 (2005).

\bibitem{Harko08} T. Harko, W. F. Choi, K. C. Wong and K. S. Cheng, JCAP \textbf{0806}, 002 (2008).

\bibitem{Mizuno} S. Mizuno, K.-I. Maeda, and K. Yamamoto, Phys. Rev. {\bf D67},  023516 (2003).
\bibitem{Wong} K. C. Wong, K. S. Cheng and T. Harko, Eur. Phys. J. \textbf{C68}, 241 (2010).
\bibitem{Lobo} F. S. N. Lobo, Phys. Rev. \textbf{D75}, 064027 (2007).

\bibitem{Leon} J. P. Leon, Mod. Phys. Lett. \textbf{A16}, 2291 (2001).

\bibitem{Ku08} P. K. F. Kuhfittig, Schol. Res. Exch. {\bf 2008}, 296158 (2008).

\bibitem{Liddle} A. R. Liddle and A. J. Smith, Phys. Rev. {\bf D68}, 061301 (2003).
\bibitem{Bowcock} P. Bowcock, C. Charmousis and R. Gregory, Class. Quantum. Grav. {\bf 17} 4745, (2000).
\bibitem{Mukohyama} S. Mukohyama, T. Shiromizu and K. Maeda, Phys. Rev. {\bf D62}, 024028 (2000).

\bibitem{Binetruy} P. Binetruy, C. Deffayet and D. Langlois, Nucl. Phys. B. {\bf 615}, 219 (2001).

\bibitem{Israel} W. Israel, Nuovo Cimento \textbf{B44}, 1 (1966).

\bibitem{Maartens} R. Maartens, D. Wands, B. A. Bassett and I. P. C. Heard,
Phys. Rev. {\bf D62}, 041301 (2000).

\bibitem{Himemoto} Y. Himemoto and T. Tanaka, Phys. Rev. \textbf{D67}, 0844014 (2003).

\bibitem{Brodie} J. H. Brodi and D. A. Easson, JCAP \textbf{0312}, 004 (2003).

\bibitem{Sami} M. Sami, N. Dadhich and T. Shiromizu, Phys. Lett. \textbf{B568}, 118 (2003).

\bibitem{Polchinski}  J. Polchinski, Phys. Rev. Lett \textbf{75}, 4724
(1995).

\bibitem{Cyburt} R. H. Cyburt, J. Ellis, B. D. Fields and K. A. Olvie, Phys. Rev. \textbf{D67}, 103521 (2003).

\bibitem{Ellis} J. R. Ellis, J. E. Kim and D. V. Nanopoulos, Phys. Lett. \textbf{B145}, 181 (1984).

\bibitem{Carr1} B. J. Carr and S. W. Hawking, Mon. Not R. Astron. Soc. \textbf{168}, 399 (1974).

\bibitem{Carr2} B. J. Carr, T. Harada and H. Maeda, Class. Quantum. Grav. \textbf{27} (2010).



\end{thebibliography}
\end{document}